# Analysis & Shortcomings of E-Recruitment Systems: Towards a Semantics-based Approach Addressing Knowledge Incompleteness and Limited Domain Coverage[1]

## Abstract

The rapid development of the Internet has led to introducing new methods for e-recruitment and human resources management. These methods aim to systematically address the limitations of conventional recruitment procedures through incorporating natural language processing tools and semantics-based methods. In this context, for a given job post, applicant resumes (usually uploaded as free-text unstructured documents in different formats such as .pdf, .doc, or .rtf) are matched/screened out using the conventional keyword-based model enriched by additional resources such as occupational categories and semantics-based techniques. Employing these techniques has proved to be effective in reducing the cost, time, and efforts required in traditional recruitment and candidate selection methods. However, the ***skill gap*** – i.e. the propensity to precisely detect and extract relevant skills in applicant resumes and job posts – and the hidden semantic dimensions encoded in applicant resumes still form a major obstacle for e-recruitment systems. This is due to the fact that resources exploited by current e-recruitment systems are obtained from generic domain-independent sources, therefore resulting in knowledge incompleteness and the lack of domain coverage. In this paper, we review state-of-the-art e-recruitment approaches and highlight recent advancements in this domain. An e-recruitment framework addressing current shortcomings through the use of multiple cooperative semantic resources, feature extraction techniques and skill relatedness measures is detailed. An instantiation of the proposed framework is proposed and an experimental validation using a real-world recruitment dataset from two employment portals demonstrates the effectiveness of the proposed approach.

## Keywords:

---

# 1. Introduction

Recently, e-recruitment platforms have become the main channels for job applicants (1, 2). These platforms have proved to be more effective than traditional recruitment methods as they provide organizations with wide geographical outreach and save time, cost, and effort required to hire the right talent (3, 4). However, current e-recruitment platforms face a major challenge with regard to the underlying techniques that they employ (5). This challenge lies in the fact that assigning relevance scores between job posts and candidate resumes is accomplished based on i) overlapping skills found in their content and ii) the exploitation of semantic resources – that suffer from knowledge incompleteness and limited domain coverage - to recognize unspecified skill entities (6, 7). Furthermore, as both employers and applicants have shifted to using online employment portals, employers started to receive large numbers of resumes that are usually uploaded as free-text unstructured e-documents in different formats such as .pdf, .doc, or .rtf (1, 2, 6).

To help employers find the right candidate from a numerous set of resumes, researchers have proposed several solutions that exploit text processing and semantics-based techniques. Examples of these techniques are skills overlap screening (8), models based on relevance feedback (9), techniques that employ the Analytic Hierarchy Processes (10), semantics-based techniques (7, 11-15), and machine learning algorithms (16-20). Although these approaches have proved to assist employers in screening out irrelevant resumes, they still suffer from low precision ratios when matching resumes to their relevant job postings (21). For instance, systems that employ text processing and highlighting of skills overlap fail to address the *skill gap* – i.e. the precise detection and extraction of skills in applicant resumes and job posts – and to detect and extract the hidden semantic dimensions encoded in applicant resumes. Consequently, results produced by these techniques are unsatisfying for employers as many of the resumes can be assumed false positives (when considering irrelevant resumes as relevant to a given job post) or false negatives (when resumes that are relevant to a given job post are not retrieved) (8). To overcome the drawbacks of traditional text and skills overlap techniques, researchers propose to utilize machine learning and feature extraction algorithms, occupational categories and classifications, and dictionaries and knowledge bases. Although employing such approaches has led to significant improvements, they still suffer from problems associated with the limited domain coverage of the exploited resources and the lack of semantic knowledge captured by such resources (22).

Inspired by the recently proposed semantics-based techniques, we present an automatic e-recruitment system that employs multiple cooperative semantic resources and occupational classifications (WordNet (23), YAGO3 (24), and Hiring Solved Dataset (25)) to screen out irrelevant resumes and precisely match candidate resumes to their relevant job posts. The proposed system starts by employing Natural Language Processing (NLP) and feature extraction techniques to convert unstructured resumes and job posts into semi-structured documents. The produced documents contain tagged-elements (we refer to these elements as *segments*) extracted from both resumes and job posts such as job experience and educational background. The documents also contain other relevant concepts extracted from the content of both resumes and job posts. These concepts are obtained based on the exploited semantic resources. When the exploited resources fail in recognizing a given concept, we utilize skill relatedness measures to compensate for such incomplete knowledge in the used semantic resources, and to further enrich the initially extracted concepts. The main contributions of our work are summarized as follows:

1. Conducting a comprehensive comparative analysis between existing e-recruitment systems and classifying them according to several categorization criteria such as the goal of the system, implementation techniques/approaches, type of input, type of output and the evaluation technique.

2. Proposing an e-recruitment approach integrating multiple semantic resources and skills relatedness techniques in an attempt to discover the hidden semantic dimensions encoded in the content of resumes and their relevant job posts. The matching process between resumes and job posts considers, unlike conventional approaches, segments of resumes with their relevant segments of job posts instead of taking into account the whole content of e-documents.

The remainder of this paper is organized as follows. Section 2 presents the related work. The general architecture of the proposed system is presented in Section 3. Section 4 discusses the detailed characterization of the proposed e-recruitment system. Section 5 introduces the results of evaluating the proposed system against other existing systems using a real-world recruitment dataset from two employment portals. In the final section, we draw the conclusions and outline future work.

## 2. Related work

In this section, we provide a review of the state-of-the-art related to e-recruitment approaches. We start with an introduction on recruitment (both conventional and automatic recruitment strategies). Then, we present the methods and techniques used by e-recruitment systems and discuss the strengths and weaknesses of these systems. Next, we present a comprehensive comparative analysis between the discussed systems and classify them according to different categorization criteria as detailed in sections 2.3 and 2.4 respectively. Finally, we summarize this chapter in section 2.5.

### 2.1 Background

As stated in (26), recruitment is defined as the process of generating a pool of job seekers whom are valuable for the company, have all necessary skills and expertise and meet all job requirements that enable them to contribute in constructing a promising future for the organization. Traditional recruitment methods are time-consuming and usually require huge efforts by HR departments. As described by the authors of (27, 28), the conventional recruitment process can be divided into the following stages:

1. Employer Branding and Applicant Attraction: this stage aims to create a good reputation for the organization in order to attract a large number of qualified applicants. To do so, employers utilize different means such as:
- Generic employment portals (e.g. Monster.com and HotJobs.com).

- Advertising about job offers such as advertising in press and on publisher websites.
- Cooperating with recruitment service providers.
2. Management: employers liaise with applicants and manage their selection process which is separated into pre-selection and selection stages.
3. Pre-selection and Candidate Selection: applicants' resumes and certificates are checked to screen out inappropriate candidates. Accordingly, applications that are not screened out during the pre-selection process are selected and evaluated in order to make final hiring decisions.

In the past, many organizations used conventional recruitment and candidate selection methods to hire employees through collecting resumes from traditional media such as newspapers, magazines, job agencies and web sites. Then, candidates are screened for choosing the most suitable persons for vacant posts. Although this recruitment and selection process performs well in screening out unqualified applicants, it still has limitations associated with the required effort, cost and time (26, 28) to match resumes to their relevant job offers. To address these issues, several e-recruitment systems have been proposed (7, 18, 20). These systems are preferred by employers and job seekers in comparison to traditional recruitment methods due to their introduced advantages (29). For example, e-recruitment systems are cost effective, easy to use, have proper targeting in any field or industry, generate fast response, allow to build up a database of candidates for talent searching, enable employers to present more information regarding the required job skills and competencies and allow them to have better access to talents (26, 29). By reviewing state-of-the-art e-recruitment systems, we find that they have employed different techniques and approaches for automating the conventional recruitment process. In the following section, we provide more details about these techniques and approaches and discuss their strengths and weaknesses.

## 2.2 E-recruitment systems: techniques and approaches

Over the past few years, researchers have proposed several systems that aim to overcome the limitations of traditional recruitment methods (1, 3, 7, 9, 10, 20, 30-33). One of the earliest fundamental techniques was based on employing NLP steps to process and further analyze the contents of both job posts and resumes. In this technique, the exact match between keywords extracted from the content of job offers and candidate resumes was the only means for deciding upon their relatedness. Systems that employed NLP techniques suffered from low precision as large portions of the automatically assigned relevance scores were not relevant. This is namely because such techniques ignore the latent semantic aspects of the contents of job offers and resumes (7).

To overcome the limitations of the traditional NLP techniques, the Structured Relevance-based Model (SRM) has been proposed (31). In this approach, relevance models (built from highly ranked documents) are used to compensate for vocabulary variations between resumes and job descriptions. Similar job offers are grouped by matching a candidate job description with a collection of job descriptions. After that, resumes that are relevant to those job descriptions are used to construct relevance models to capture terms that are not explicitly mentioned in job descriptions. A major problem of this approach lies in its low precision when tested against large-scale real-world datasets (31).

Other researchers have studied the impact of using occupational classifications and additional semantic resources on improving the precision of e-recruitment systems. As stated in (14), the exploitation of semantic resources in the recruitment domain assists in using shared vocabularies to describe job descriptions and resumes. The authors of (12, 13, 15, 34) propose exploiting occupational classifications and/or semantic resources that have been built based on integrated classifications and standards. In (15), the authors use a human resource ontology (HR-ontology) to gain uniform representation of resumes and job offers and to accomplish the matching process at the semantics level. Another semantics-based system is EXPERT (7) which constructs ontology documents that describe both job offers and resumes based on the concept linking approach (35), and then ontology documents of job offers are mapped to ontology documents of resumes. The authors of (33) propose exploiting a set of manually-constructed description logic based concept lattices and filters to tag resumes and job offers with semantic descriptors. These descriptors are used in the matching process to identify qualified and over qualified candidate resumes. In this work, the authors argue that sophisticated knowledge bases in the HR domain are still rare. Accordingly, the authors propose building a knowledge-based representation of job offers and user profiles using manually-constructed rules and filters. According to the authors, new rules can be defined to extend the initially constructed concept lattices. In the same manner as proposed in this work, the authors of (32) propose using Formal Concept Analysis (FCA) to extend and maintain existing ontologies that are exploited in the HR domain. The authors argue that using high-level general ontologies for the purpose of matching resumes and job offers is not effective. This is because such ontologies are generic and lack coverage of various domain-specific recruitment concepts. To address this issue the authors propose using FCA for updating existing ontology hierarchies with new subsumption relations or new concepts.

Although these approaches have shown better results in accomplishing the matching process, they still face significant problems concerned with the development of complete and reliable ontologies that capture up-to-date knowledge about specific domains (22). Furthermore, we show that instead exploiting conventional approaches and algorithms for building and enriching existing ontological hierarchies, knowledge captured in existing semantic resources and occupational categories can be integrated to cooperatively assign relevance scores between resumes and job offers. More details on this approach are provided in Section 4.

## 2.3 Classification of existing e-recruitment systems

In this section, we present a comparative analysis between existing e-recruitment systems and classify them according to the following categorization criteria:

- **Goal of the system:** as will be further discussed in this section, the reviewed systems have two main goals. They either aim to find a strict match between job posts and resumes (i.e. Boolean model) or they focus on ranking applicants' resumes according to their relevance to a given job post. In the context of the second type of system, employers can detect whether an applicant is under qualified, qualified or even over qualified for a given job offer.

- **Implementation techniques/approaches**: to classify e-recruitment systems we also consider the techniques/approaches that are employed by each system. These techniques include keyword-based screening, semantics and occupational category based methods, machine learning algorithms, and a combination of these approaches.
- **Type of input**: e-recruitment systems accept different types of input. The input (resumes and job posts) can be in the form of structured (using forms), semi-structured (using xml generated document), or unstructured (in .pdf or .doc format) documents. In the context of the research work, we are mainly concerned with unstructured e-documents which are the most challenging to consider.
- **Type of output**: another important criterion that we consider for categorizing e-recruitment systems is the type of output that each system produces. Basically, the output produced by e-recruitment systems can belong to one of two categories. In the first category, the produced results are characterized by their relevance/non relevance to a given job post. The systems of the second category extend this approach by producing ranked results. In this context, such systems do not only filter a given set of resumes (i.e. match/ not match), but they also recommend highly ranked resumes to their relevant job posts.
- **Testing and evaluation method**: different evaluation mechanisms have been carried out to test and evaluate the effectiveness of the proposed recruitment systems, and to find whether the returned results (resumes) by each system are true positives (i.e. relevant to a given job post and were retrieved by the system). To do this, researchers have conducted experiments using real-world recruitment scenarios and manually-crafted datasets, while others have implemented system prototypes wherein they tested the overall effectiveness of the employed techniques. We would like to point out that evaluating the techniques and approaches employed in e-recruitment systems is of great interest as they can be successfully adopted in practical settings and have their positive impact on the revenue models of the companies that adopt them.

In the rest of this section, we discuss different e-recruitment systems, describe their characteristics and classify them according to the introduced set of categorization criteria.

### 2.3.1 The impact of semantic web technologies on job recruitment processes

This system is one of the earliest systems that exploited semantic resources to find matches between job offers and their corresponding resumes (13). The authors exploit a human resource ontology (also referred to as semantic resource) - constructed by integrating widespread standards and classifications - to annotate the content of job offers and resumes. In order to collect candidate resumes, web-based application forms are used to acquire CVs as semi-structured resumes. Then, the human resource ontology is utilized to detect the semantic aspects of the produced semi-structured resumes and job posts. Finally, a semantic matching algorithm is employed to generate a list of qualified applicants. However, although semantics-based approaches enhance the effectiveness of e-recruitment systems (15), they are penalized by limitations of the exploited semantic resources, namely semantic knowledge incompleteness and limited domain coverage (22). On the other side, in the proposed approach, the authors rely on web-based application forms to acquire CVs as semi-structured resumes. This would be a tedious and time-consuming task for applicants (2).

### 2.3.2 EXPERT

In (7), the authors propose to match between resumes and job posts based on employing semantics and knowledge-based methods similarly to the previously mentioned system. However, in order to start the matching process, this system first produces ontological representations of resumes and job posts to detect knowledge encoded in their contents. After that, the ontology documents (ontological representations) of resumes are mapped to ontology documents of job offers to retrieve relevant candidates. In this context, an ontology mapping (36) approach is utilized to determine the correspondences between the concepts of the produced ontology documents. To measure the effectiveness of the proposed system, the authors evaluate its precision in assigning relevance scores between job offers and applicant resumes. In order to accomplish this task, two CV sets are used. The first CV set consists of structured resumes while the second CV set consists of unstructured resumes and job posts. The results show high precision and recall ratios indicating the effectiveness of employing semantics and knowledge based methods in the domain of e-recruitment. Nevertheless, when we compare this system with the system proposed in (6), we find that the latter has been more effective and precise in matching resumes to job posts.

### 2.3.3 On-line consistent ranking on e-recruitment: seeking the truth behind a well-formed CV

In the work presented in (5), job applications are evaluated and ranked by exploiting semantics-based matching techniques and machine learning algorithms. First, the proposed system extracts a set of features from the applicants' LinkedIn profiles and matches them semantically against job posts. In order to accomplish this task, a single semantic resource has been constructed by domain experts to derive the semantic aspects of resumes and job offers. In addition, linguistic analysis is utilized to analyze candidates' blogs to extract features that reflect their personality traits and social behaviors. Afterwards, supervised machine learning algorithms are used to generate a list of qualified applicants ranked according to their relevance. Although employing machine learning and semantics-based techniques have proved to assist employers in screening out irrelevant resumes, they still suffer from limitations, namely semantic knowledge incompleteness and limited domain coverage stemming from the resources (training data, ontologies and knowledge bases). This system is evaluated in a real-world recruitment scenario by comparing manually calculated scores between resumes and job posts to those produced by the system. The results have shown acceptable accuracy except for job offers that require special skills.

### 2.3.4 MatchingSem

MatchingSem (37) is an e-recruitment system that matches unstructured documents (resumes and job posts) based on employing multiple semantic resources and statistical-based techniques. The proposed system first employs NLP tools to find and extract lists of candidate concepts from the content of both resumes and job offers. Next, existing semantic resources are employed to analyze the lists of candidate concepts at the semantics level. When a concept is not recognized by the used semantic resources, statistical-based concept-relatedness measures are then used to address this issue. To evaluate the effectiveness of the methods and techniques employed in the proposed system, an experimental instantiation is conducted by comparing manually assigned scores between resumes and job posts and those produced by the proposed system in the same manner as carried out in (5). Although the system shows high precision and recall ratios for most of the examined job posts, its overall performance is hindered by the *skill gap* as inferior precision and recall results are exhibited for job posts that require specific skills in terms of years of experience.

### 2.3.5 Matching Resumes and Jobs based on Relevance Models

This system has been proposed to match semi-structured resumes and job offers in real-world large scale recruitment scenarios (31). It as well supports applicants ranking according to their similarity scores. Relevance models are built from known relevant resumes to a specific job post and used to compensate for vocabulary variations between resumes and job descriptions. Similar job offers are grouped by matching a candidate job description with a collection of job descriptions. Afterwards, resumes that are relevant to those job descriptions are used to construct relevance models to capture terms that are not explicitly mentioned in job descriptions. A major problem of this approach is its low precision when tested in large-scale real-world datasets.

### 2.3.6 E-Gen

E-Gen (9, 17) is an automatic e-recruitment system that matches unstructured resumes to their relevant job posts. It is based on employing Support Vector Machine (SVM) classification algorithms in order to annotate segments of job offers with the appropriate topics and features. Additionally, E-Gen addresses the issue of ranking applicants according to their relevance score by utilizing the vector space model. In this context, job offers and resumes are transformed into vector space representations and then similarity measures for their associated vectors are computed. Relevance feedback is then utilized to expand the job post vector representation with terms extracted from relevant candidate resumes. Next, similarity measures are recomputed in order to ameliorate the produced results. An experimental instantiation of the proposed system is conducted to prove its effectiveness in a real-world recruitment scenario. However, the utilized SVM classification algorithms are subjective to high error rates since they depend on manually developed training corpora (1).

### 2.3.7 Application of machine learning algorithms to an e-recruitment system

In this approach (18), an e-recruitment system is proposed based on a machine learning paradigm. It starts by analyzing job posts and semi-structured resumes acquired by web-based forms and applicants' LinkedIn profiles. Then, machine learning algorithms are utilized to produce a list of qualified applicants ranked according to their relevance. In this context, the ranking process mainly focuses on learning a scoring function that calculates relevance scores between resumes and their relevant job posts. Therefore, a set of training data is collected by domain experts to further learn the required scoring function. An experimental instantiation of the proposed system has been installed to validate its effectiveness in strictly matching resumes against job posts. Although the authors argue that the produced results are satisfying in identifying applicant's personality traits, the consistency of the produced results (i.e. lists of qualified applicants ranked according to their relevance) is highly dependent on the job posts. For example, it is difficult to learn a scoring function for senior positions which require specific experience and skills.

### 2.3.8 Convex

Convex (38) is an automatic e-recruitment system built to match unstructured/semi-structured resumes to job posts. The proposed system starts by employing a single domain-specific knowledge base in an attempt to extract concepts from both job posts and candidate resumes. If the used knowledge base fails in identifying a specific concept, extraction techniques are then utilized to compensate for missing background knowledge. Concept extraction techniques include shallow natural language parsing and heuristics. On the one hand, shallow natural language parsing uses two domain-independent, language-specific NLP techniques to extract noun phrases as concepts (i.e. barrier word algorithm and parts-of-speech tagging). On the other hand, rule-based heuristics are employed by domain experts to further extract other relevant concepts that were not captured by NLP techniques. Once concepts are extracted, the matching process produces a list of qualified applicants. To validate the effectiveness of the proposed system, the authors compare manually assigned relevance scores between resumes and job posts with those produced automatically by the proposed system. The results show that Convex performs better than approaches employing keyword-based or statistical-based techniques. However, the proposed system is penalized by the use of a single generic knowledge base. Accordingly, if the concepts found in resumes and job posts are not captured by the used knowledge base due to its limited coverage then the system fails drastically in finding relevant applicants.

### 2.3.9 A hybrid approach to managing job offers and candidates

The system (1) is an extended version of the E-Gen system that utilizes a hybrid approach combining statistical-based algorithms and vector space representations to match resumes and job posts. The proposed system appends a summarization module to exclude irrelevant information contained in resumes and cover letters according to specific compression criteria determined by employers. The updated version of the system deals with: i) the extraction of information from job posts, ii) the processing of resumes and cover letters, iii) the computation of relevance scores between resumes and job posts. According to the authors, in order to evaluate the precision of the proposed system, experimental validations are carried out on a dataset consisting of 1917 resumes and 3 job posts. Although the produced results are satisfactory, the proposed system is hindered by limitations associated with the summarization module wherein resumes and cover letters are subjective to excluded relevant information.

## 2.4 Comparative analysis between e-recruitment systems

As shown in Table 1, we have conducted a comparative analysis between existing e-recruitment systems/approaches and classified them according to different categorization criteria. We can see that most of the above mentioned systems focus on matching resumes with job posts while a few of them additionally attempt at ranking applicants according to their relevance scores. On the other hand, the type of input varies from one e-recruitment system to another. Some systems accept unstructured resumes and job offers as input, while others are concerned with structured or semi-structured resumes and job offers. We would like to point out that – in the context of our work – we aim to analyze and match unstructured resumes to job posts as they are the most trivial and common form of submitted e-documents. Concerning the employed techniques and approaches, it is clear that semantics-based techniques and machine learning algorithms are the dominant techniques and have been exploited by most of the systems. This is due to the fact that semantic resources play a crucial role in attempting to highlight the semantic aspects hidden in the content of both resumes and job offers.

**Table 1.** Classification of the studied e-recruitment systems

| Index | System | Goal | Technique | Type of input | Type of output | Testing and evaluation method |
|---|---|---|---|---|---|---|
| 2.3.1 | The Impact of Semantic Web Technologies on Job Recruitment Processes | Matching resumes to job offers | Semantics-based technique | Semi-structured resumes and job offers | List of candidate applicants | A prototypical implementation of the system without using experiments in real-world scenario |
| 2.3.2 | EXPERT | Matching resumes to job posts | Semantics-based technique | Structured / unstructured resumes and job posts | List of candidate applicants | Evaluated using two data sets of Structured and unstructured resumes and job posts |
| 2.3.3 | On-line Consistent Ranking on E-recruitment: Seeking the Truth Behind a Well-Formed CV | Matching resumes and Ranking applicants | Semantics and machine learning algorithms | Structured resumes and job posts | List of applicants ranked according to their relevance scores | Evaluated in real-world recruitment scenario. The system shows good accuracy except for job posts that require special skills |
| 2.3.4 | Matchingsem | Matching resumes to job posts | Semantics and statistical based techniques | Unstructured resumes and job posts | List of candidate applicants | Evaluated in real-world recruitment scenario. The system shows good accuracy except for job posts that require special years of experience |
| 2.3.5 | Matching Resumes and Jobs Based on Relevance Models | Matching resumes and Ranking applicants | Structured Relevance Models | Semi-structured resumes and job posts | List of candidate applicants | Evaluated in a large-scale real-world recruitment scenario by comparing manually assigned scores and those produced by the |

| | | | | | system |
|---|---|---|---|---|---|
| 2.3.6 | E-Gen | Matching resumes and Ranking applicants | Machine learning and vector space model | Unstructured job posts and resumes / cover letters | List of applicants ranked according to their relevance scores | Evaluated in real-world recruitment scenario |
| 2.3.7 | Application of Machine Learning Algorithms to an Online Recruitment System | Matching resumes and Ranking applicants | Machine learning and linguistic analysis | Structured / unstructured resumes and job posts | List of applicants ranked according to their relevance scores | Evaluated in real-world recruitment scenario. The results show that the system is effective in identifying personality traits |
| 2.3.8 | A Hybrid Approach to Managing Job Posts and Candidates | Matching resumes and Ranking applicants | Machine learning algorithms, statistical-based techniques and vector space model | Unstructured job posts and resumes / cover letters | List of applicants ranked according to their relevance scores | Evaluated in real-world recruitment scenario using a huge dataset resumes |
| 2.3.9 | Convex | Matching resumes to job posts | Semantics and shallow natural language processing | Unstructured / semi-structured resumes and job posts | List of candidate applicants | Evaluated in real-world recruitment scenario by comparing manually assigned scores to those produced by the system |

Considering the testing and evaluation methods of the studied systems and approaches, we can notice that some experiments do not bring to light the precision of the evaluated systems since they do not rely on a significant real-world recruitment scenario. To judge the quality of results generated from these systems, manually assigned relevance scores (a.k.a. expert judgments or ground truth) are usually compared to their corresponding automatically generated matching scores.

# 3. General architecture of the proposed system

In this section, we present a general overview of the proposed e-recruitment system wherein multiple cooperative semantic resources and statistical-based skills relatedness measures are combined to effectively match between job posts and their relevant resumes. Figure 1 depicts the overall architecture of the proposed system which is comprised of several processing modules organized as follows:

- When applicants and employers upload their resumes and job offers (in the form of unstructured .doc, .pdf, or. rtf files), the first module entitled **From Unstructured Documents to Semi-Structured Documents** converts the received unstructured files into semi-structured documents. This step is important as instead of matching unstructured versions of resumes and job offers, the system matches tagged-segments of resumes to their relevant segments in the job offers. More elaboration on this step will be presented in Sec 4.1.

The next module of the system is the ***Concept Identification and Extraction*** module. This module is employed to detect candidate matching concepts from the content of the semi-structured versions of the job postings and resumes. To carry out this step, the system utilizes various Natural Language Processing (NLP) tools such as n-gram tokenization, stop words removal, and Part-of-Speech Tagging (POST).

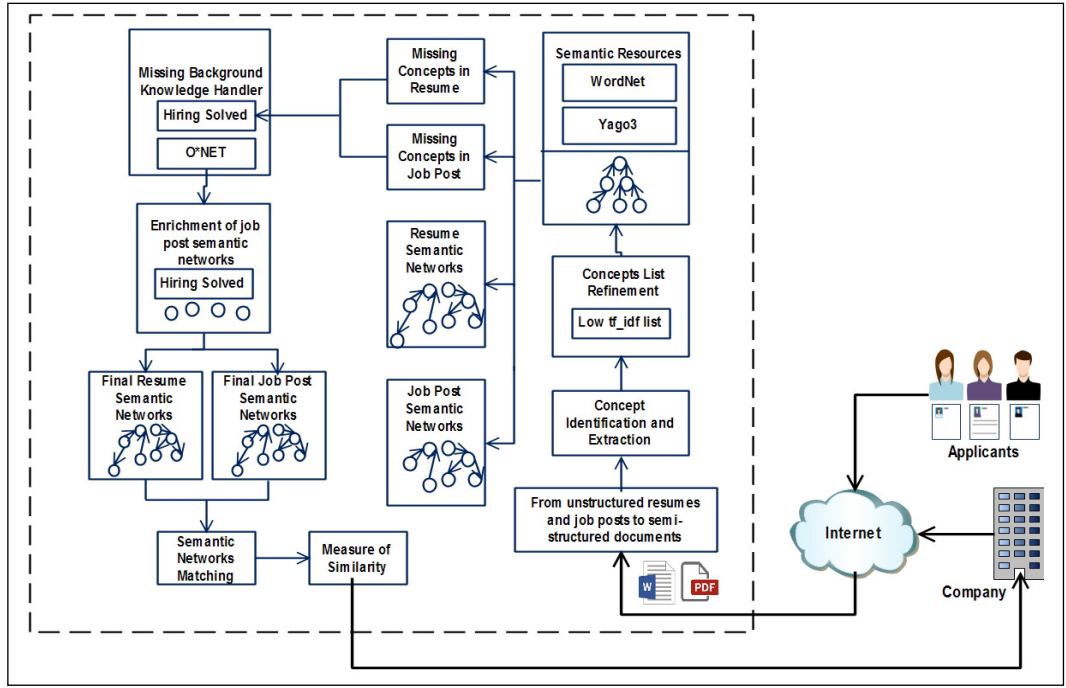

Figure 1. General architecture of the proposed system

- By employing the ***Refinement of Candidate Concepts*** module, the system removes concepts that have little contribution in the matching procedure. Examples of these concepts are those that usually fall under specific sections in the resume such as: candidate's name, address, contact information, etc. In addition, concepts that have low tf-idf weights (8) are removed as detailed in section 4.2.

- The refined lists of concepts (from the segments of both the job offers and resumes) are then submitted to the exploited ontologies (WordNet and YAGO3) to construct semantic networks (wherein concepts are connected by various types of semantic relations). Details of these semantic resources are listed below.

1. WordNet (23): a generic lexical database created manually to cover different domains. It groups the concepts into sets of synonyms called synsets. These synsets are connected with different types of semantic relations such as hypernymy, meronymy and hyponymy. WordNet is primarily used for automatic text analysis and word sense disambiguation. Additionally, we utilize it in our system to discover semantic relations among different concepts in resumes and job posts.

2. YAGO3 (24): a large high quality semantic resource developed at the Max Planck Institute for Computer Science in Saarbrücken. YAGO3 is an extension of the YAGO knowledge base that combines the information from the Wikipedia in multiple languages. It is manually evaluated by finding the correctness of 4412 facts. 98.07% of the evaluations were judged to be correct.

The number of semantic networks that may be produced at this step can vary from one to multiple semantic networks. All of the produced networks from the resume segments will be matched to their corresponding networks that are extracted from the job offer. It is important to point out that – due to the lack of domain coverage by the exploited ontologies - some concepts are not recognized by the exploited semantic resources. To handle this issue, we utilize the ***Missing Background Knowledge Handler*** where additional occupational categories such as Hiring Solved (HS) dataset (25) and O*NET are used to enrich the constructed semantic networks with additional semantically-related concepts. Both categories define a large number of terms in the form of skills – either mentioned in job offers or resumes – and the weights of the semantic relatedness between those skills. In this context, semantically-relevant concepts are extracted and used to expand the constructed semantic networks. It is important to point out that we have manually enriched these datasets with new concepts to achieve broader domain coverage.

- In the semantic networks matching module, the updated semantic networks are regarded as input to the matching algorithm. The algorithm produces measures of semantic relatedness between the networks that are derived from the segments of each resume to their corresponding networks that are derived from a given job post. As a result, each resume is ranked according to the total weight that is assigned to it by the matching algorithm.

# 4. Detailed characterization of the proposed e-recruitment system

Before we detail the proposed system architecture, we formalize the use of the terms "Semantic Resource", "Semantic Network", "Semantic Network Enrichment" and ''tf-idf weighting''.

**Definition 1: Semantic Resource:**

A semantic resource $\Omega$ is a quintuple $<C, R, I, V, A>$ where:

- $C$ is the set of classes (i.e. concepts) defined in the semantic resource. The class hierarchy of $\Omega$ is a pair $(C, \leq)$ where $\leq$ is an order relation on $C \, x \, C$. We call $\leq$ the sub-class relation.

- $R$ is the set of relations.

- $I$ is the set of individual that are used to represent instances of the semantic resource classes.

- $V$ is the set of relation values.

- *A* is the set of axioms (such as constraints).

In the proposed approach, the system regards the lists of candidate concepts (obtained from job posts and resumes) as input, and creates as output two sets of semantic networks. These are:

- The set of semantic networks $S_j$ that are constructed based on the job posts.

- The set of semantic networks $S_r$ that are constructed based on the resumes.

These semantic networks are automatically constructed based on the exploited semantic resources.

Formally, we define a semantic network as follows.

**Definition 2: Semantic Network:**

A semantic network $\zeta$ is defined as a triplet *<C, R, A>* where:

- *C* is the set of concepts identified in the semantic network. These concepts represent the semantic aspects of resumes and job posts.

- *R* is set of relations derived from the exploited semantic resources that hold between concepts of *C*.

- *A* is the set of axioms defined on *C* and *R* according to $\Omega$.

As highlighted in the previous section, despite the fact that we are employing multiple semantic resources, we may find that some concepts are not recognized due to the issue of knowledge incompleteness and limited domain coverage. To overcome this issue, we exploit Hiring Solved Dataset to enrich the semantic networks of both job posts and resumes. Formally, we define the process of semantic network enrichment as follows:

**Definition 3: Enrichment of Semantic Network:**

It is defined as the process that takes a given semantic resource $\Omega$ and a given concept *c* as input and produces for *c* a set $E(c) \subseteq C_\Omega$ as output where *E(c)* is the set of suggested enrichment candidates for *c*.

**Definition 4: Tf-idf weighting:**

The *tf-idf* weighting scheme [9] produces a weight $W_c$ for a concept **c** in a document **d** according to:

$$W_c = tf_{c,d} * idf_c \tag{1}$$

where $tf_{c,d}$ is the frequency of concept *c* in document *d* (e.g. the number of times that *c* occurs in *d*) and the inverse document frequency $idf_c$ is a measure of the degree of informativeness of *c*, i.e. whether it appears frequently or not in all the considered documents.

It is important to mention that we employ the *tf-idf* weighting scheme at the corpus level in order to eliminate the concepts that have no significant meaning among the set of candidate concepts $S_c$ – obtained using the NLP pre-processing techniques detailed in

section 4.2. The set of relevant concepts $S_r$ is obtained based on a threshold value $t$ according to:

$$S_r = \{c \in S_c \mid W_c \leq t\} \tag{2}$$

## 4.1 From unstructured resumes and job posts to semi-structured documents

In this first processing module, unstructured resumes and job posts (that are uploaded as .doc or .pdf files) are converted into semi-structured documents based on employing the Apache Tika toolkit[2] for accessing the documents, as well as other feature extraction techniques. These techniques include regular expressions and the following NLP steps:

1. Resume/job post segmentation: the content of resumes/job posts is divided into units (paragraphs or sentences) and then each unit is processed separately.
2. N-gram tokenization: each unit is split into unigram, bigram and trigram tokens. At this step, n-grams are submitted to the exploited semantic resources (WordNet and YAGO3) to capture compound terms, as well as their synonyms. For example, when submitting the term "software engineer" to WordNet, we will obtain a list of synonyms (highlighted using Bold font style) to this term, in addition to other hypernyms (highlighted using Italics and Underline font styles) as shown below:

---

**Sense of software engineer**

Sense 1
**programmer**, **computer programmer**, **coder**, **software engineer** -- (a person who designs and writes and tests computer programs)
    => _engineer_, _applied scientist_, _technologist_ -- (a person who uses scientific knowledge to solve practical problems)
    => _computer user_ -- (a person who uses computers for work or entertainment or communication or business)

---

We submitting the same term to YAGO3, it redirects the system to find the synonyms of the term "Programmer" in WordNet. This happens as YAGO3 integrates concepts defined in WordNet. We would like to point out that for other compound terms, YAGO3 may return additional semantically relevant terms. In this context, the system refines the set of the extracted concepts by enriching them with those that are obtained from the used semantic resources. It is also important to mention that many of the n-grams, namely tri-grams are missing in WordNet, and this was due to the fact that it has a very limited domain coverage compared to YAGO3 which comprises millions of entities and facts about those entities. For example, the tri-gram "software development lifecycle" is missing in WordNet, however, YAGO3 recognized this concept and returned the below semantically-relevant terms:

---

```
<Software_Engineering_Institute>
<Software_development>
<Business_rule_management_system>
<Structured_analysis>
<Software_quality_management>
<Microsoft_Visual_Studio>
<Runtime_intelligence>
<Schema_migration>
<Development_testing>
<Telerik>
```

It is important to point out that some acronyms may be missing (not recognized) in the exploited semantic resources. For example, the acronym "jsp", is not recognized neither by WordNet nor by YAGO3. To address this issue, we utilize the missing background knowledge handle that is detailed in section 4.4.

3. Stop word removal: a list of words that have no semantic significance is defined. Then, these words are removed to enhance the system performance.

4. Part-of-speech tagging: each token is assigned its part-of-speech category such as noun, verb, adverb, etc.

5. Named Entity Recognition (NER): it seeks to classify tokens into a set of predefined categories such as person, duration, number and location. In the context of our work, we define rules for labeling tokens with "DEGREE", "EDUCATION FIELD" and "EXPERIENCE" entities through using RegexNER Stanford CoreNLP[3]. Examples of these rules are shown below.

| | |
|---|---|
| *Bachelor of (Arts\|Laws\|Science\|Engineering)* | *DEGREE* |
| *PhD* | *DEGREE* |
| *Master of (Arts\|Laws\|Science\|Engineering)* | *DEGREE* |
| *M.Sc.* | *DEGREE* |
| *B.Sc.* | *DEGREE* |
| *Information Technology* | |
| *EDUCATION FIELD* | |
| *CS* | *EDUCATION FIELD* |
| *Computer Science* | *EDUCATION FIELD* |
| *Software engineer* | *EXPERIENCE* |
| *Java programming language* | *EXPERIENCE* |

The following example clarifies the process of converting unstructured resumes and job posts to semi-structured documents. It is important to point out that the used job posts and resumes are real resumes that have been collected from various online portals such as (https://www.indeed.com/resumes?isid=find-resumes&ikw=hometop&co=US&hl=en).), as well as from university academic staff members. The below example demonstrates the details of the above-mentioned steps:

**Example 1**: Converting unstructured resume and job post to semi-structured documents.

- Part of example job post (P1):

---

> *If you are a Java Developer with experience, please read on.*
> *We move quickly and innovate constantly to deliver exciting online game experiences to players around the world.*
> **What you will be doing**
> - *Design, develop and maintain backend systems written in Java and/or Node.js.*
> - *Identify scaling bottlenecks and propose solutions.*
> - *Work in close partnership with a team of diverse and talented peers in various disciplines including design, development, operations, PM's and SDET's for sustained long term success. Partner with the architects and the technical leadership team to deliver solid technical designs.*
> - *Participate actively in detailed design, code reviews, bug/issue triage with the feature teams, and support.*
>
> **What you need for this position**
> - *3+ years of experience in Java programming language (e.g. jsp)*
> - *Bachelor of Science in Computer Science.*
> - *You should be a programmer who is looking to take his experience to the next level.*

- Part of example applicant resume (CV1):

> **Java developer**
> ***Personal summary***
> A skilled java developer with proven expertise in using new tools and technical developments to drive improvements throughout an entire software development lifecycle. Having extensive industry and full life cycle experience in a java based environment, along with exceptional analytical, design and problem-solving capabilities. Excellent communication skills and able to work alongside support teams and the java community to define and refine new functionality.
> Looking for ambitious company which will challenge my developer and problem solving skills and allow me to continue to develop my knowledge and potential.
> ***Key skills and experience***
> Strong core Java, j2ee, jsp, xml development experience.
> Ability to develop creative solutions for complex problems.
> I have worked as a Software engineer for 2 years.
> ***Education***
> B.Sc. in CS.
> M.Sc. in CS.

We convert segments of P1 and CV1 from unstructured documents to semi-structured documents as follows:

| Semi-structured document of CV1 | Semi-structured document of P1 |
|---|---|
| *<Job post Info>*<br>*<Experience>*<br>*<Years>3</Years>*<br>*<Field>Java programming language </Field>*<br>*</Experience>*<br>*<Education>*<br>*<Degree> Bachelor of Science</Degree>*<br>*<Field> Computer Science</Field>*<br>*</Education>*<br>*</Job post Info>* | *<Applicant Info>*<br>*<Experience>*<br>*<Years>2</Years>*<br>*<Field>Software engineer</Field>*<br>*</Experience>*<br>*<Education>*<br>*<Degree> B.Sc.</Degree>*<br>*<Field>CS</Field>*<br>*</Education>*<br>*<Education>*<br>*<Degree> M.Sc.</Degree>*<br>*<Field>CS</Field>*<br>*</Education>*<br>*</Applicant Info>* |

We first apply regular expressions to identify the job experience section (paragraph or sentence). Some of these regular expressions are shown below.

```
1:  [0-9]+(\\-[0-9]+)?\\+? years .+ experience
2:  .+? months .+ experience
3:  work.+ .+ years
```

Then, NLP techniques are performed to extract the number of years of experience (tokens that are labeled as "NUMBER" or "DURATION") and experience field (tokens that are labeled as "EXPERIENCE"). After that, we identify educational background info such as education degree (tokens that are labeled as "DEGREE") and education field (tokens that are labeled as "*EDUCATION FIELD*").

## 4.2 Concept extraction and refinement

In this second processing module, candidate concept lists of resumes and job posts are extracted and identified based on executing NLP steps that have been mentioned in section 4.1. When these steps are performed, lists of concepts that represent both the job post and resume are identified. The next example clarifies the process of concept extraction from P1 and CV1 based on NLP steps.

**Example 2**: Concept extraction.

In this example, we consider a part (i.e. one segment) of both the job post (P1) and the resume (CV1) due to space restrictions as shown below.

- The Segment of job post (P1):

  **What You Need for this Position**

  *3+ years of experience in Java programming language (e.g. jsp)*

  *Bachelor of Science in Computer Science.*

  *You should be a programmer who is looking to take his experience to the next level.*

- The segment of resume (CV1):

  **Key skills and experience**

*Strong core Java, j2ee, jsp, xml development experience.*

*Ability to develop creative solutions for complex problems.*

*I have worked as a Software engineer for 2 years.*

***Education***

*B.Sc. in CS*

*MSc in CS*

The content of the resume/job post is first divided into paragraphs/sentences and then each paragraph/sentence is processed separately. Afterwards, n-gram tokenization is performed and stop words are removed according to a predefined list of words such as: *a*, *the*, *we*, *his*, *(*, *)*, *is*. Then, the word category disambiguation and the NER steps are carried out using the StanfordCoreNLP. In the context of our work, nouns (NNP, NNPS, and NN) are included in the lists of candidate concepts. The results of applying these steps are shown in Table 2.

**T**able 2 Results of applying the NLP steps

| Candidate concepts extracted from the segment of job post (P1) | Candidate concepts extracted from the segment of resume (CV1) |
|---|---|
| Programmer | Core |
| experience | development |
| Java | experience |
| programming language | Java |
| Jsp | j2ee |
|  | Jsp |
|  | Xml |
|  | software engineer |
|  | ability |

After the extraction process, the tf-idf weighting scheme is utilized to identify concepts that have no significant meaning and may negatively impact the matching process. Accordingly, those that have low tf-idf weights are removed from the lists of candidate concepts.

## 4.3 Construction of semantic networks

In this section, we detail the process of constructing semantic networks that represent the lists of refined candidate concepts and the semi-structured documents. Each concept is submitted to WordNet ontology (23) in order to extract the semantic and taxonomic relations (synonymy relation – referred to as "same as" - and hypernymy relation – referred to as "is a" -) that hold with other concepts. Figure 2 depicts the output of the module performing the construction of semantic networks on our example.

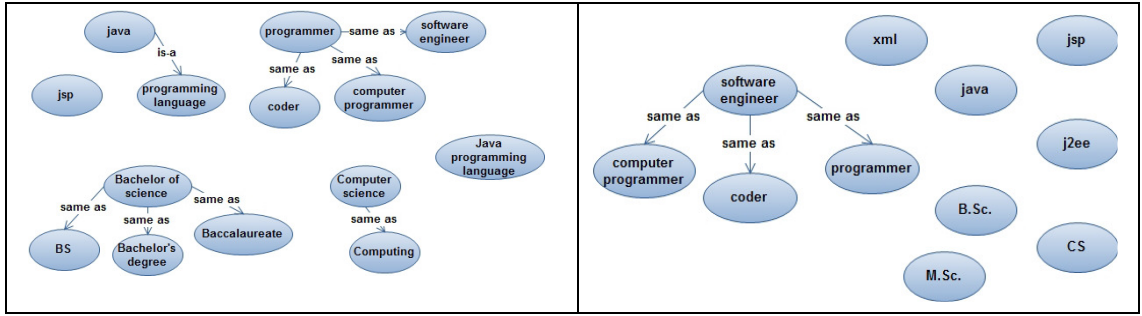

Figure 2: a) Semantic networks obtained from the job post (P1)   b) Semantic networks obtained from the resume (CV1)

When we explore the hierarchy of WordNet ontology, we can see that the term "java" which exists in the example job post (P1) and example resume (CV1) has three different senses (i.e. meanings):

1. Java -- (an island in Indonesia south of Borneo; one of the world's most densely populated regions)
2. Coffee, java -- (a beverage consisting of an infusion of ground coffee beans; "he ordered a cup of coffee")
3. Java -- (a simple platform-independent object-oriented programming language used for writing applets that are downloaded from the World Wide Web by a client and run on the client's machine).

It is clear that in the context of both P1 and CV1, "java" refers particularly to the third sense. Therefore, we employ a Word Sense Disambiguation (WSD) technique to specify the correct sense for each term according to its surrounding textual content. To do this, we have utilized the list of extracted concepts (that belong to the segment in the job offer where the original term appeared) to detect the correct sense of a given term. In this context, when we have a term with multiple senses such as "java", the definition of each sense is tokenized in the same manner as we do for tokenizing the text segment of this term (referred to as $L_{java}$ in our example) in the job offer. However, we do not consider synonyms of the tokens and only identify compound terms. In addition, we apply the same stop words removal step to remove stop words from the definition of each sense. Next, we find the similarity scores between the concepts that belong to the text segments of each term's sense and the concept that belong to the paragraph that the original term appeared at. In the current version of the system, the highest similarity score is considered for judging the relevance between a given term and its senses. For example, when tokenizing the definition of each sense of the term "java" we will get the following concept lists:

1. L1: java, island, indonesia, south, borneo, one, world, densely, populate, region
2. L2: coffee, java, beverage, infusion, ground, coffee bean, order, cup
3. L3: java, programming, programming language, object-oriented programming language, write, applet, download, world wide web, client, run, machine

Based on the above extracted lists of concepts, the similarity scores are computed by utilizing the jaro-winkler distance function detailed in section 4.6. We used this distance function to mitigate the problem of superficial differences (such as hyphenated terms, plurals and compounds) between each input term from Li and those that belong the L_{java} segment.

Besides, the synonyms of each disambiguated terms are used to expand the constructed semantic networks. The rest of concepts that are missing from the WordNet ontology are then submitted to the YAGO3 ontology. Accordingly, semantic relations that are defined in YAGO3 are also exploited to expand the constructed semantic networks. However, we would like to point out that even using a second ontology like YAGO3 may not fully address the missing background knowledge problem since some concepts such as "jsp" are not defined in it. Therefore, concepts that are not recognized in the WordNet or YAGO3 ontologies are submitted to the missing background knowledge handler in section 4.4. We would like to point out that the constructed semantic networks represent ad-hoc ontologies that are developed based on utilizing the exploited semantic resources, as well as the newly obtained concepts from HD dataset.

## 4.4 Missing background knowledge handler

When the exploited semantic resources fail to recognize a given concept from the lists of refined candidate concepts, the Hiring Solved (HS) dataset (39) is then employed to compensate for such missing background knowledge. This knowledge source defines a large number of terms in the form of skills – either mentioned is job posts or resumes – and the weights of semantic relatedness between them. For example, although the term "jsp" was not recognized by the exploited semantic resources, when we submit it to the HS dataset we get a set of semantically relevant terms to "jsp" as shown in Table 3. The weights shown in Table 3 represent measures of semantic relatedness between the submitted term and its related terms.

**Table 3** The result of submitting "jsp" to HS dataset

| Term | Relatedness Measure |
| --- | --- |
| servlets | 1.00 |
| j2ee | 0.94 |
| Jdbc | 0.92 |
| tomcat | 0.90 |
| Ejb | 0.76 |
| struts | 0.75 |
| hibernate | 0.62 |
| xml | 0.60 |
| java | 0.56 |

Following this step, concepts in semi-structured documents that are missing in the used semantic resources are submitted to O*NET in order to recognize concepts tagged in the produced semi-structured documents which are not fully covered in the used semantic resources. It is important to point out that we have manually enriched this dataset with missing concepts to ensure broader domain coverage. A subset of this data set is shown in Table 4.

**Table 4** Subset of manually defined dataset by domain expert

| Term | Relation | Term |
|------|----------|------|
| B.Sc. | same as | Bachelor of science, BSc, B.Sc, BS, Bachelors, Bachelor,.B.S. |
| M.Sc. | same as | Master of science, MSc, M.Sc, Master"s degree |
| CS | same as | Computer Science |
| SE | same as | Software Enginering |
| CSE | same as | Computer System Engineering |
| IT | same as | Information Technology |
| Computer Science | related to | SE, CSE,IT |
| Computer Network Architect | related to | Network Analyst, Network Consultant, Network Engineer, Network Manager, Networking Systems and Distributed Systems Engineer, Systems Engineer, Telecommunications Analyst, Telecommunications Engineer |

Based on the results of applying the missing background knowledge handler, the semantic networks are updated as depicted in Figures 3 and 4.

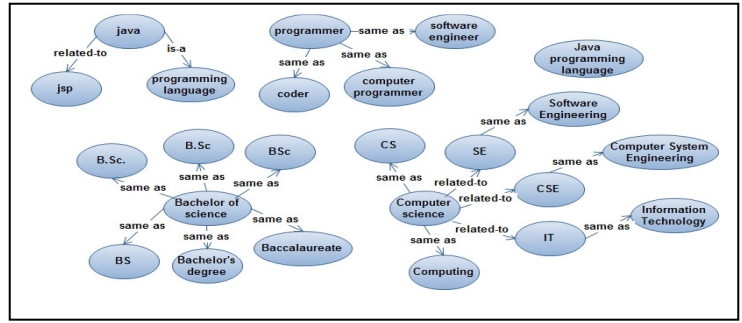

Figure 3. Updated semantic networks built from the example job post P1

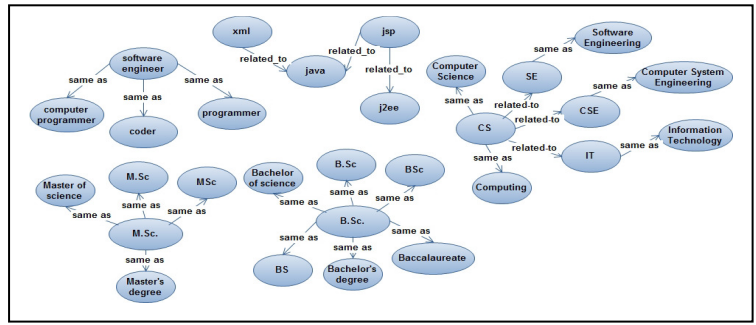

Figure 4. Updated semantic networks built from the example resume CV1

As shown in Figures 3 and 4, concepts in the semantic networks built from P1 and CV1 are connected with the newly obtained concepts from the HS dataset and the manually constructed dataset. For instance, we can see that the numerical degree of semantic relatedness between the terms "j2ee" and "jsp" is 0.94. We replace this semantic relatedness values by the *related-to* relation and use it to connect both concepts.

## 4.5 Further enrichment of the produced semantic networks

Semantic networks constructed from job posts represent the reference to which semantic networks generated from resumes are matched. In this context and since some of the required skills may not be explicitly defined by the employer, we further enrich the semantic networks of the job posts by automatically adding new skills obtained from HS dataset. To carry out this step, we submit the job titles to HS dataset to obtain a set of related skills to each title. For instance, when submitting the job title (i.e. "java programmer") of job post (P1) to HS dataset, it returns the list of skills shown in Figure 5. As highlighted in the previous section, we replaced the measures of semantic relatedness with the *related-to* relation and only considered the top 5 related skills returned.

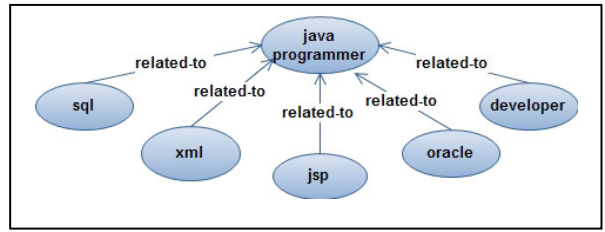

Figure 5. Top 5 skills related to the title "Java Programmer" returned by HS dataset

To enrich the semantic networks constructed from a job post *p* with a candidate concept *c*, we follow the following procedure:

- If *c* already exists in the semantic networks built from *p*, then we retain *c* in its position in the networks. For example, since the element "jsp" is already defined in the semantic networks built from P1, we keep this element in its position in the network.

- If *c* does not exist in the semantic networks built from *p*, then we update the networks by adding the job title as a new node and then attaching it to all other candidate concepts that do not exist in the semantic networks built from *p*. Figure 6 shows the enrichment of the semantic networks built from our example job post P1.

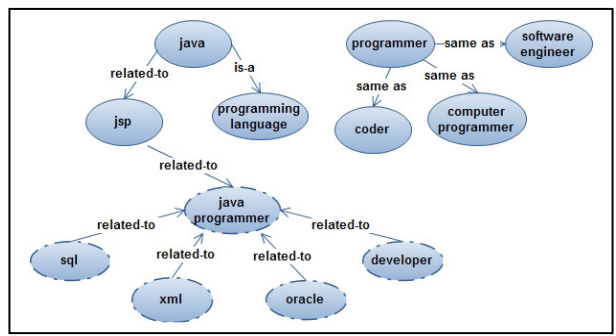

Figure 6. Enrichment of the semantic networks built from the job post P1

# 4.6 Matching of semantic networks

During the matching process, we use a multi-level matching algorithm to match between the semantic networks built from resumes and job posts. Firstly, we match the semantic networks that represent the acquired/required "educational background information". Secondly, we match the semantic networks that represent job experience information in both resumes and job posts. And finally, we match the semantic networks of candidate concepts. In this context, we use Algorithm 1 to match between the semantic networks as described below.

---

**Algorithm 1. Name-based technique for finding the similarity between the resume semantic network ($SN_R$) and the job post semantic network ($SN_J$)**

**Input**: $SN_R$ and $SN_J$

**Output**: Measure of similarity based on correspondences set $S$

1: *answer* ← ⟨ ⟩ ;

2: for i←0; i < $SN_J$ .Length; i++

3:    for j←0; j< $SN_R$ .Length; j++

4:       *answer* ← **JWinkler**($SN_J$ [i] $SN_R$ [j])

5:       if(*answer* < v) then

6:          add(SN_J[i], SN_R[j]) to $S$

7:       end if

8:    end for

9: end for

10: **return** *similarity*

---

Algorithm 1 above is employed to find the similarity between the resume semantic network ($SN_R$) and the job post semantic network ($SN_J$). Each run of the algorithm these semantic networks are regarded as input respectively: $SN_R$ and $SN_J$ derived from the "educational background information" segment, $SN_R$ and $SN_J$ derived from the "job experience information" segment, and $SN_R$ and $SN_J$ that are derived from the "candidate concepts" list. To do this, the JWinkler (known as Jaro-Winkler (40)) distance function is used. This function is a simple and fast technique that measures the similarity between the strings of concepts and instances in both networks. The Jaro-Winkler similarity metric between two string $s$ and $t$ is given by:

$$jaro - winkler(s,t) = \frac{1}{3} * \left\{ \frac{m}{|s|} + \frac{m}{|t|} + \frac{m-t^{`}}{m} \right\} \tag{3}$$

where:

- $s$ : is the first string
- $t$ : is the second string
- $m$ : is the number of matching characters
- $t^{`}$ : is the number of transpositions

This algorithm produces as output a correspondences set S. This set includes common concepts between the semantic networks of resumes and job posts and it is further used to find relevance scores between each resume and its relevant job post based on Equation 4. This equation is an adapted form of the candidate's relevance scoring (RS) formula that has been proposed in the Oracle Project Resource Management (Management, 2010). The formula for calculating the scoring percentage is as follows:

$$Rs = \frac{|\{S_{cc}\}|}{|\{C_{cj}\}|} * 70\% + \frac{|S_{e}|}{|\{C_{ej}\}|} * 15\% + \frac{|\{S_{x}\}|}{|\{C_{xj}\}|} * 15\% \tag{4}$$

Where:

- $S_{cc}$ : the correspondences set of candidate concepts.
- $C_{cj}$ : the candidate concepts of the job post.
- $S_{e}$ : the correspondences set of concepts that describe educational background information.
- $C_{ej}$ : the concepts that represent educational background information in the job post.
- $S_{x}$ : the correspondences set of concepts that describe job experience information.
- $C_{xj}$ : the concepts that represent experience information in the job post.

It is important to point out that, the weighting values are variable and can be determined according to the employers' preferences. In the context of our work, we have assigned the following weighting values:

- Candidate concepts weight = 70%.
- Educational level weight = 15%.
- Job experience weight = 15%.

We would like to highlight that although the weighting values are variable, we have decided upon using the above mentioned values since they are the actual values that have been manually assigned during the phase of constructing our testing ground truth. This accordingly ensures conducting a fair evaluation of the effectiveness of the proposed system (i.e. when comparing the automatically generated relevance scores by the system to the manually assigned scores).

## 5. Experimental results

In the following sections, we discuss experiments in terms of two different aspects. First, we discuss the experiments that we carried out to compare between relevance scores that are produced by our system when utilizing feature extraction techniques against when not using them. Second, we experimentally demonstrate the effectiveness of the proposed system in assigning relevance scores between job posts and their relevant resumes. We implemented all solutions in Java and experiments were performed on a PC with dual-core CPU (2.1GHz) and (4 GB) RAM. The operating system is Windows 10.

In our recruitment scenario, we collected a data set of 500 resumes downloaded from http://www.amrood.com/resumelisting/listallresume.htm and other local job portals, and used ten job posts that have been obtained from http://jobs.monster.com. The collected resumes and job posts are unstructured documents in different document formats such as (.pdf) and (.doc). In order to carry out the experiments, we converted unstructured resumes and job posts to semi-structured documents by highlighting the educational background and job experience information in order to enhance the matching process. After that, we analyzed the corpus of the resumes and job posts through employing NLP techniques as described in section 4.1. Then, we utilized multiple semantic resources to construct the semantic networks of resumes and job posts such as WordNet and YAGO3.

Additionally, the constructed networks for job posts were further enriched based on HS dataset to add additional skills that are not explicitly defined by the employer. And finally, the resulting networks were automatically matched and different relevance scores were produced by the system.

## 5.1 A comparison between the system effectiveness when utilizing feature extraction techniques against not utilizing them.

In this section, we compare between the produced results by the system when we utilize feature extraction (FE) techniques to extract the experience and educational background information against when not utilizing them. In order to provide a ground for evaluating the quality of the produced results, we manually calculated all relevance scores (RS) between each job post and its relevant resumes. Then, we compared the manually calculated scores to those produced by the system when considering feature extraction techniques and when only using candidate concepts identification modules.

As shown in Table 5, we have three job posts, and for each job posts we have six resumes. The first job post requires a java developer with the following characteristics: 5+ years of server side design and development experience, B.Sc. degree in Computer Science and knowledge in object oriented programming language such as (Java, C++), REST based web service development and http principles. The second job post requires 6+ years of professional experience related to system testing, Bachelor's degree in computer science or related field and knowledge and experience with tracking and testing tools such as Selenium, SoapUI, Remedy and Siebel. The third job post centers on looking for talented candidates with 6 years of software engineering experience responsibilities such as agile/iterative development methodologies (XP, SCRUM, etc.), object-oriented design and Java programming skills. As we can see in Table 5, the manual scores that were assigned for each resume by our expert are very close to those produced by the system when utilizing feature extraction techniques. For example, if we consider the first job post (Java developer) and the fourth resume (IT-testing that describes an applicant with Bachelor of Computer Applications degree (B.C.A) and 2.5 years of software testing experience), we can see that the difference between the manually assigned score and the automatically generated score when utilizing FE techniques is less than when not utilizing them. This is due to integrating two new factors in calculating RS as shown in equation 4.

Accordingly, automatic RS between IT-testing and Java developer is increased by 0.15 due to a match between the acquired and required educational background info. However, for some particular results, integrating FE techniques doesn't affect the produced relevance score. For example, when we consider the second job post (i.e. senior test engineer) and the third resume (i.e. IT-CRM that describes an applicant with master of computer applications degree and 2.2 years of experience in Client / Server based applications development and support), we can see that the automatic RS equals the manually assigned RS. This is due to the fact that there is no match between the required and acquired educational background and job experience info. And hence, capturing the experience and education information from the resume and job post doesn't affect (i.e. increase) the automatic RS.

**Table 5** The system results using/not using IE module

| Job post | Resumes | Manual Relevance Scores | Automatic Relevance Scores using Feature Extraction Techniques | Automatic Relevance Scores without Feature Extraction Techniques |
|---|---|---|---|---|
| | IT-QA | 0.38 | 0.45 | 0.30 |
| | Software engineer Fresh graduate | 0.26 | 0.19 | 0.04 |
| **Java Developer** | IT-CRM | 0.15 | 0.18 | 0.18 |
| | IT-Programming Not exciplicitly mentioned | 0.3 | 0.36 | 0.21 |
| | IT-testing | 0.3 | 0.29 | 0.14 |
| | Network admin | 0.3 | 0.27 | 0.27 |
| | IT-QA | 0.45 | 0.46 | 0.31 |
| | Software engineer | 0.15 | 0.22 | 0.07 |
| **Senior test engineer** | IT-CRM | 0.1 | 0.11 | 0.11 |
| | IT-Programming | 0.25 | 0.26 | 0.11 |
| | IT-testing | 0.46 | 0.5 | 0.35 |
| | Network admin | 0.1 | 0.19 | 0.19 |
| | IT-QA | 0.61 | 0.66 | 0.36 |
| | Software engineer | 0.37 | 0.25 | 0.10 |
| **Software engineer** | IT-CRM | 0.15 | 0.15 | 0.15 |
| | IT-Programming | 0.38 | 0.40 | 0.25 |
| | IT-testing | 0.38 | 0.43 | 0.28 |
| | Network admin | 0.3 | 0.23 | 0.23 |

## 5.2 Experiments using expert judgments

In this section, we evaluate the system effectiveness based-on comparing the manually assigned relevance scores between resumes and their related job posts and automatically generated scores. In this context, we used the Precision (P) indicator in order to measure the quality of our results. This measure is defined as follows:

**Precision (P)**: is the Percentage Difference between the automatically assigned relevance scores (between each job post and its relevant resumes) and those automatically generated by the system.

$$P = \frac{|V_{manual} - V_{automatic}|}{(\frac{V_{manual} + V_{automatic}}{2})} * 100\% \qquad (4)$$

where:

- $V_{manual}$: is the manually assigned relevance score between each resume and job post.

- $V_{automatic}$: is the automatically calculated relevance score between each resume and job post.

**Table 6** Precision results using IE module

| Job post | Resumes | Manual Score | Automatic Score | Precision (%) |
|---|---|---|---|---|
| **Java developer** | IT-Mobile | 0.22 | 0.18 | 0.80 |
| | IT-Systems | 0.23 | 0.27 | 0.84 |
| | Electronic eng | 0.10 | 0.16 | 0.54 |
| | IT_prog | 0.30 | 0.30 | 1.00 |
| **Senior test engineer** | IT-Mobile | 0.10 | 0.07 | 0.65 |
| | IT-Systems | 0.10 | 0.16 | 0.54 |
| | Electronic eng | 0.10 | 0.03 | 0.54 |
| | IT_prog | 0.25 | 0.27 | 0.46 |
| **Database developer** | IT-Mobile | 0.10 | 0.07 | 0.54 |
| | IT-Systems | 0.23 | 0.28 | 0.81 |
| | Electronic eng | 0.10 | 0.18 | 0.43 |
| | IT_prog | 0.24 | 0.23 | 0.96 |

As shown in Table 6, the manual scores that were assigned for each resume by our expert are very close to the automatically calculated scores by the system. This is because we have integrated two new important factors (educational background and job experience info) in calculating relevance scores. These factors constitute 30% of the final result (relevance score). In addition, we have employed multiple semantic resources and statistical concept-relatedness measures to represent the semantic aspects of resumes and job posts and to further enrich them with concepts that are not recognized by the used semantic resources.

However, we can find that for some particular results the percentage difference was large. For example, when matching the second job post "senior test engineer" and "Electronic eng" resume, the difference is (0.46 i.e. 100% - 54%). This is because the job post has optional requirements in its job description such as (having knowledge and experience with tracking tools such as Remedy, Siebel, or other industry standard). This optional requirement is not distinguished from other obligatory requirements by our system and thus the manual score for the resume is larger than the automatic score. In order to solve this problem, we plan to assign different weights for optional and obligatory requirements, and then use these weights in computing the relevance scores between job posts and resumes.

# 6. Conclusions

In this paper, we have conducted a comprehensive analysis of existing e-recruitment systems and categorized them according to a set of evaluation criteria. In addition, we have presented our proposed solution and detailed its implementation steps. Unlike conventional e-recruitment systems, the proposed system exploited multiple semantic resources such as WordNet and YAGO3, as well as other NLP, feature extraction and skills relatedness techniques in an attempt to discover the hidden semantic dimensions encoded in the content of resumes and their relevant job posts. The matching process between resumes and job offers considers, unlike conventional approaches, segments of resumes with their relevant segments of job posts instead of taking into account the whole content of e-documents. In addition, we have utilized HS dataset to address the issues of missing background knowledge in the exploited semantic resources on the one hand, and to enrich job posts with further semantically relevant concepts on the other. Initial experiments using a real-world dataset that comprises resumes and job posts that belong to different domains showed promising precision results and proved the effectiveness of the employed techniques in assigning relevance scores between candidate resumes and their corresponding job offers. However, it is important to point out that there are still a number of limitations in the current version of the system. One of these limitations is the complexity of the matching algorithm due to the utilization of several semantic resources that consist of millions of entities such as YAGO3. The complexity of the algorithm will also increase with each addition of a new sematic resource. To address this issue, we plan to construct an integrated semantic resource that comprises several resources. In this context, and instead of routing resumes/job posts to each semantic resources independently, they will be routed towards a single integrated resource. Another weakness in the proposed approach lies in the fact that each new resume will be matched with all of the offered job posts in the system. This means that we have a huge search space (a very large pool of job offers) that needs to be accessed at each matching step. To address this limitation, we plan to build an updated version of the system wherein we will employ classification techniques to classify job offers according to the occupational categories that they belong to. In this context, for each new resume, the system will match it with the job offers that cover the occupational category/ies that the resume belongs to. Accordingly, we aim to achieve two main benefits. On the one hand, the matching space will be minimized, and on the other hand the run-time complexity of the matching procedure will be reduced. In the future work, we also plan to testify the proposed system using additional resumes and job posts. In addition, we plan to propose a job offer recommender module, where applicants will receive automatic job recommendations based on the analysis of their resumes. We plan to test the impact of the proposed recommendation module on the applications and extend it to recommend candidate resumes to employers who are seeking qualified applicants that meet their job requirements.

# References

1.      Kessler M, B N, Chet, Roche M, Torres-Moreno J-M, El-B M, et al. A hybrid approach to managing job offers and candidates. Inf Process Manage. 2012;48(6):1124-35.

2.      Brandão C, Morais C, Dias S, Silva AR, Mário R, editors. Using Online Recruitment: Implicit Theories and Candidates' Profile. World Conference on Information Systems and Technologies; 2017: Springer.

3.      Singh A, Rose C, Visweswariah K, Chenthamarakshan V, Kambhatla N. PROSPECT: a system for screening candidates for recruitment.  Proceedings of the 19th ACM international conference on Information and knowledge management; Toronto, ON, Canada. 1871523: ACM; 2010. p. 659-68.

4.      Javed F, Hoang P, Mahoney T, McNair M, editors. Large-Scale Occupational Skills Normalization for Online Recruitment2017.

5.      Faliagka E, Iliadis L, Karydis I, Rigou M, Sioutas S, Tsakalidis A, et al. On-line consistent ranking on e-recruitment: seeking the truth behind a well-formed CV. Artificial Intelligence Review. 2014;42(3):515-28.

6.      Kmail AB, Maree M, Belkhatir M, Alhashmi SM. An Automatic Online Recruitment System based on Multiple Semantic Resources and Concept-relatedness measures. In Proceedings of the 23rd IEEE International Conference on Tools with Artificial Intelligence (ICTAI'15). 2015.

7.      Kumaran VS, Sankar A. Towards an automated system for intelligent screening of candidates for recruitment using ontology mapping EXPERT. Int J Metadata Semant Ontologies. 2013;8(1):56-64.

8.      Belkin NJ, Croft WB. Information filtering and information retrieval: two sides of the same coin? Commun ACM. 1992;35(12):29-38.

9.      Kessler R, Béchet N, Torres-Moreno J-M, Roche M, El-Bèze M. Job Offer Management: How Improve the Ranking of Candidates. In: Rauch J, Raś Z, Berka P, Elomaa T, editors. Foundations of Intelligent Systems. Lecture Notes in Computer Science. 5722: Springer Berlin Heidelberg; 2009. p. 431-41.

10.     Faliagka E, Ramantas K, Tsakalidis AK, Viennas M, Kafeza E, Tzimas G, editors. An Integrated e-Recruitment System for CV Ranking based on AHP. WEBIST; 2011.

11.     Colucci S, Di Noia T, Di Sciascio E, Donini FM, Mongiello M, Mottola M. A formal approach to ontology-based semantic match of skills descriptions. J UCS. 2003;9(12):1437-54.

12.     Trichet F, Bourse M, Leclere M, Morin E, editors. Human resource management and semantic Web technologies. Information and Communication Technologies: From Theory to Applications, 2004 Proceedings 2004 International Conference on; 2004 19-23 April 2004.

13.     Bizer C, Heese R, Mochol M, Oldakowski R, Tolksdorf R, Eckstein R. The impact of semantic web technologies on job recruitment processes. Wirtschaftsinformatik 2005: Physica-Verlag HD; 2005. p. 1367-81.

14.     Mochol M, Paslaru E, Simperl B. Practical guidelines for building semantic erecruitment applications. International Conference on Knowledge Management, Special Track: Advanced Semantic Technologies (AST'06). 2006:1-8.

15.     Mochol M, Wache H, Nixon L. Improving the Accuracy of Job Search with Semantic Techniques. In: Abramowicz W, editor. Business Information Systems. Lecture Notes in Computer Science. 4439: Springer Berlin Heidelberg; 2007. p. 301-13.

16.     Chung-Kwan S, Ui Tak Y, Huy Kang K, Sang Chan P. A hybrid approach of neural network and memory-based learning to data mining. Neural Networks, IEEE Transactions on. 2000;11(3):637-46.

17.     Kessler m, Torres-Moreno JM, El-B M, ze. E-Gen: automatic job offer processing system for human resources.  Proceedings of the artificial intelligence 6th Mexican international conference on Advances in artificial intelligence; Aguascalientes, Mexico. 1776068: Springer-Verlag; 2007. p. 985-95.

18.     Faliagka E, Ramantas K, Tsakalidis A, Tzimas G, editors. Application of machine learning algorithms to an online recruitment system. ICIW 2012, The Seventh International Conference on Internet and Web Applications and Services; 2012.

19.     Faliagka E, Tsakalidis A, Tzimas G. An integrated e-recruitment system for automated personality mining and applicant ranking. Internet research. 2012;22(5):551-68.

20.     Hong W, Zheng S, Wang H, Shi J. A Job Recommender System Based on User Clustering2013.

21.     Suerdem A, Akalin M. Using Conjoint Analysis to Determine the Requirements of Different Users for Designing Online Solution Tools: Job Matching Platform. In: Bilgin MH, Danis H, Demir E, Lau CKM, editors. Innovation, Finance, and the Economy. Eurasian Studies in Business and Economics. 1: Springer International Publishing; 2015. p. 239-52.

22.     Maree M, Belkhatir M. Addressing semantic heterogeneity through multiple knowledge base assisted merging of domain-specific ontologies. Knowl-Based Syst. 2015;73:199-211.

23.     Miller GA. WordNet: a lexical database for English. Commun ACM. 1995;38(11):39-41.

24.     Mahdisoltani F, Biega J, Suchanek FM. YAGO3: A Knowledge Base from Multilingual Wikipedias. 2015.

25.     HiringSolved. HiringSolved website 2015. Available from: https://hiringsolved.com/explorer.

26.     Sivabalan L, Rashad Y, Nor Haslinda I. How to Transform the Traditional Way of Recruitment into Online System: International Business Research;Mar2014, Vol. 7 Issue 3; 2014.

27.     Lang S, Laumer S, Maier C, Eckhardt A, editors. Drivers, challenges and consequences of E-recruiting: a literature review. Proceedings of the 49th SIGMIS annual conference on Computer personnel research; 2011: ACM.

28.     Färber F, Weitzel T, Keim T. An automated recommendation approach to selection in personnel recruitment. AMCIS 2003 Proceedings. 2003:302.

29.     Pande S. E‐recruitment creates order out of chaos at SAT Telecom. Human Resource Management International Digest. 2011;19(3):21-3.

30.     Lee I. An architecture for a next-generation holistic e-recruiting system. Commun ACM. 2007;50(7):81-5.

31.     Yi X, Allan J, Croft WB. Matching resumes and jobs based on relevance models. Proceedings of the 30th annual international ACM SIGIR conference on Research and development in information retrieval; Amsterdam, The Netherlands. 1277920: ACM; 2007. p. 809-10.

32.     Looser D, Ma H, Schewe K-D, editors. Using formal concept analysis for ontology maintenance in human resource recruitment. Proceedings of the Ninth Asia-Pacific Conference on Conceptual Modelling-Volume 143; 2013: Australian Computer Society, Inc.

33.     Paoletti AL, Martinez-Gil J, Schewe K-D, editors. Extending Knowledge-Based Profile Matching in the Human Resources Domain2015; Cham: Springer International Publishing.

34.     García-Sánchez F, Martínez-Béjar R, Contreras L, Fernández-Breis JT, Castellanos-Nieves D. An ontology-based intelligent system for recruitment. Expert Systems with Applications. 2006;31(2):248-63.

35.     Senthil Kumaran V, Sankar A. Expert locator using concept linking. International Journal of Computational Systems Engineering. 2012;1(1):42-9.

36.     KALFOGLOU Y, SCHORLEMMER M. Ontology mapping: the state of the art. The Knowledge Engineering Review. 2003;18(01):1-31.

37.     Aseel B. Kmail, Mohammed Maree, Belkhatir M. MatchingSem: Online Recruitment System based on Multiple Semantic Resources.  12th IEEE International Conference on Fuzzy Systems and Knowledge Discovery (FSKD'15)2015.

38.     Dan C, editor A Hybrid Approach to Concept Extraction and Recognition-Based Matching in the Domain of Human Resources2004.

39.     Liu P, Azimi J, Zhang R, editors. Automatic keywords generation for contextual advertising. Proceedings of the 23rd International Conference on World Wide Web; 2014: ACM.

40.     Winkler WE, editor The state of record linkage and current research problems. Statistical Research Division, US Census Bureau; 1999: Citeseer.